\documentclass[epj,nopacs]{svjour}
\usepackage{graphicx} 
\usepackage{amsmath}
\usepackage{bm,amsfonts,amssymb}
\usepackage{color} 
\newcommand{\be}{\begin{equation}}
\newcommand{\ee}{\end{equation}}
\newcommand{\bdm}{\begin{displaymath}}
\newcommand{\edm}{\end{displaymath}}

\newcommand{\barr}{\begin{array}}
\newcommand{\earr}{\end{array}}

\newcommand{\tso}{${}^{3\!}S_1$}
\newcommand{\tdo}{${}^{3\!}D_1$}
\newcommand{\tpz}{${}^{3\!}P_0$}

\newcommand{\otpo}{$1\,{}^{3\!}P_1$}
\newcommand{\ttpo}{$2\,{}^{3\!}P_1$}

\newcommand{\jpsi}{J/\psi}

\newcommand{\dzdsz}{D^0D^{\star0}}
\newcommand{\dzdbsz}{D^0\bar{D}^{\star0}}
\newcommand{\dcdsc}{D^\pm D^{\star\mp}}
\DeclareMathAlphabet{\mathitbf}{OML}{cmm}{b}{it}
\begin{document}
\title{Unquenched quark-model calculation of $\bm{X(3872)}$ electromagnetic
decays}
\author{Marco Cardoso\inst{1} \and George Rupp\inst{2} \and
 Eef van Beveren\inst{3} 
}                     
\institute{
Centro de F\'{\i}sica Te\'{o}rica de Part\'{\i}culas,
Instituto Superior T\'{e}cnico, Universidade de Lisboa,
P-1049-001 Lisbon, Portugal
\and
Centro de F\'{\i}sica das Interac\c{c}\~{o}es Fundamentais,
Instituto Superior T\'{e}cnico, Universidade de Lisboa,
P-1049-001 Lisbon, Portugal
\and
Centro de F\'{\i}sica Computacional,
Departamento de F\'{\i}sica, Universidade de Coimbra,
P-3004-516 Coimbra, Portugal
}
\date{Received: date / Revised version: date}
%
\abstract{
A recent quark-model description of $X(3872)$ as an unquenched
$2\,{}^{3\!}P_1$ $c\bar{c}$ state is generalised by now including all relevant
meson-meson configurations, in order to calculate the widths of the
experimentally observed electromagnetic decays $X(3872)\to\gamma\jpsi$ and
$X(3872)\to\gamma\psi(2S)$. Interestingly, the inclusion of additional
two-meson channels, most importantly $D^\pm D^{\star\mp}$, leads to a sizeable
increase of the $c\bar{c}$ probability in the total wave function, although the
$D^0\bar{D}^{\star0}$ component remains the dominant one. As for the
electromagnetic decays, unquenching strongly reduces the $\gamma\psi(2S)$
decay rate yet even more sharply enhances the $\gamma\jpsi$ rate, resulting
in a decay ratio compatible with one experimental observation but in slight
disagreement with two others. Nevertheless, the results show a dramatic
improvement as compared to a quenched calculation with the same confinement
force and parameters. Concretely, we obtain
$\Gamma\left(X(3872)\to\gamma\psi(2S)\right)=28.9$~keV
and $\Gamma\left(X(3872)\to\gamma\jpsi\right)=24.7$~keV, with branching ratio
$\mathcal{R}_{\gamma\psi}=1.17$.
}   


\maketitle

\section{Introduction}
\label{intro}
Since its discovery \cite{PRL91p262001} by the Belle Collaboration in 2003, 
the very narrow axial-vector \cite{PDG2014} charmonium state $X(3872)$ has
become one of the favourite theoretical laboratories for meson spectroscopists,
because of its remarkable closeness to the $D^0\bar{D}^{\star0}$
(or $\bar{D}^0D^{\star0}$) and 
$\rho^0\jpsi$ thresholds, besides its seemingly too low mass for mainstream
quark models. The now established \cite{PRL110p222001} $J^{PC}=1^{++}$ quantum
numbers seem to imply that $X(3872)$ is either the still unconfirmed
\cite{PDG2014} $\chi^\prime_{c1}$ ($2\,{}^{3\!}P_1$ $c\bar{c}$) meson, or an
axial-vector charmonium-like state of a different kind. For a recent review,
see e.g.\ Ref.~\cite{PPNP67p390}.
 
However, in order to understand the true nature of $X(3872)$, one can
ignore neither the presence of relatively nearby $1^{++}$ states in the
theoretical charmonium spectrum, nor their strong coupling to the $S$-wave
threshold $\dzdbsz$. In this spirit, the properties of $X(3872)$ were recently
studied in Refs.~\cite{EPJC71p1762,EPJC73p2351}, by modelling it as an
unquenched $\chi^\prime_{c1}$ state with additional meson-meson (MM)
components, most importantly $D^0D^{\star0}$.\footnote{For notational
simplicity, we henceforth omit the bars over the anticharm mesons.}
In the former paper \cite{EPJC71p1762}, a momentum-space calculation of
$X(3872)$ was carried out, employing the Resonance-Spectrum Expansion (RSE),
with all relevant two-meson channels included. This work showed that the
hadronic decays of $X(3872)$ can thus be described quite accurately, dispensing
with \em ad hoc \em \/tetraquark or molecular approaches. On the other hand,
the latter paper \cite{EPJC73p2351} focused on the $X(3872)$ wave function,
using instead a coordinate-space model and with only two channels, viz.\
$c\bar{c}$ and $D^0D^{\star0}$. The purpose was to study whether
the charm-anticharm component would remain substantial, despite the very
long tail of the $D^0D^{\star0}$ component due to the small binding of less
than 0.2~MeV \cite{PDG2014}. Indeed, a $c\bar{c}$ probability of about
7.5\% was found and ---
even more importantly --- a corresponding wave-function component in the inner
region of the same order of magnitude as that of the $\dzdsz$ channel, thus
ruling out a pure molecular scenario for $X(3872)$. Similar interpretations of
$X(3872)$ were concluded in the unquenched model calculations of
Refs.~\cite{PRL105p102002} and \cite{PRC88p015207}.

Besides the mentioned hadronic decays, $X(3872)$ has also been observed
\cite{PDG2014} to decay in electromagnetic (EM) processes, namely to
$\gamma\jpsi$ and
$\gamma\psi(2S)$. Such decays are very sensitive to details of the $X(3872)$
wave function, especially in its inner region, and so may discriminate among
different microscopic models.
Thus, the coordi\-nate-space method for unquenched quarkonium states employed
in Ref.~\cite{EPJC73p2351} appears to be the indicated approach for such a
calculation. Now, it was shown in Refs.~\cite{PRD81p014029,PRD86p113007} that,
in a multichannel system with one almost unbound
channel, more strongly bound channels should not be neglected beforehand for
processes in which the wave function at short distances is important. This is
of course all the more true for the $\dcdsc$ channel in the $X(3872)$
case, which is bound by only 8~MeV and so is expected to have a significant
effect on the interior wave functions of the other components. Moreover,
this channel contains charged mesons, which will contribute directly to 
EM transition amplitudes. Nevertheless, for completeness we also include all
other OZI-allowed channels with combinations of pseudoscalar and/or
vector charm-light as well as charm-strange mesons. As for the wave
functions of the EM decay products $\jpsi$ and $\psi(2S)$, again all channels
with pairs of ground-state $D$, $D^\star$, $D_s$, and $D_s^\star$ mesons will
be accounted for, since they all may develop non-negligible components.

In the present paper, we shall closely follow the formalism for
EM decays of unquenched, ``unitarised'' quarkonium systems as
developed in Ref.~\cite{PRD44p2803}. The organisation is as follows.
In Sec.~\ref{model}, a multichannel Schr\"{o}dinger equation for confined
$q\bar{q}$ channels coupled to free MM channels is written down and
solved analytically.
Section~\ref{charmonium} is devoted to the computation and display of the
multicomponent wave functions of the charmonium states $X(3872)$, $\jpsi$,
and $\psi(2S)$, using the generic solutions derived in Sec.~\ref{model}.
In Sec.~\ref{em} the procedure \cite{PRD44p2803} for EM decays of quarkonium
states with MM components in the wave function is reviewed.
Section~\ref{results}
presents the results for the EM decays, in comparison with several other
published model calculations. A summary and some conclusions are given
in Sec.~\ref{conclusions}.

\section{Multichannel Schr\"{o}dinger model}
\label{model}
Just as in Ref.~\cite{EPJC73p2351}, we consider $X(3872)$ a coupled
charm-anticharm and MM system.  The transitions between the
$c\bar{c}$ and MM sectors are assumed to take place via \tpz\ $q\bar{q}$
creation/annihilation at a sharp distance $a$, thus mimicking string breaking. 
Such a transition potential can be described in coordinate space via a
spherical delta function. This choice makes the coupled-channel equations
analytically solvable, provided that the confining $c\bar{c}$ potential allows
solutions in terms of known functions that can be analytically continued. Thus,
like in Refs.~\cite{EPJC73p2351,EPJC71p1762}, we take a harmonic oscillator
(HO) with universal frequency $\omega$. The present generalisation beyond the
model of Ref.~\cite{EPJC73p2351} amounts to the inclusion of several MM
channels instead of only one.

The coupled-channel Schr\"{o}dinger equation to be solved reads
\begin{equation}
\left[\begin{array}{cc}
\hat{h}_{q\bar{q}}^{c} & V_{cj}\\
V_{jc}^{\dagger} & \hat{h}_{MM}^{j}
\end{array}\right]\left[\begin{array}{c}
u_{c}\\
v_{j}
\end{array}\right]=E\left[\begin{array}{c}
u_{c}\\
v_{j}
\end{array}\right] \; ,
\label{eq:schro}
\end{equation}
where $\hat{h}_{q\bar{q}}$ is the quark-antiquark Hamiltonian with a
confining HO potential given by
\begin{equation}
\hat{h}_{q\bar{q}}^{c}=m_{q}^{c}+m_{\bar{q}}^{c}+\frac{\hbar^{2}}{2\mu_{c}}\left(-\frac{d^{2}}{dr^{2}}+\frac{l_{c}(l_{c}+1)}{r^{2}}\right)+\frac{1}{2}\mu_{c}\omega^{2}r^{2} \; ,
\end{equation}
with $\mu_c$ the reduced quark mass
$m_{q}^{c}m_{\bar{q}}^{c}/(m_{q}^{c}+m_{\bar{q}}^{c})$,
and $\hat{h}_{MM}$ is the free MM Hamiltonian
\begin{equation}
\hat{h}_{MM}^{j}=M_{1}^{j}+M_{2}^{j}+\frac{\hbar^{2}}{2\mu_{j}}\left(-\frac{d^{2}}{dr^{2}}+\frac{L_{j}(L_{j}+1)}{r^{2}}\right) \; .
\end{equation}
Furthermore,  $V_{cj}$ in Eq.~(\ref{eq:schro}) is the transition potential
modelled through a spherical delta shell with radius $a$
\begin{equation}
V_{cj}=\frac{\lambda g_{cj}}{2\mu_{c}}\delta(r-a) \; ,
\end{equation}
where $g_{cj}$ is the relative coupling constant of the confined $q\bar{q}$
channel $c$ to the $j$-th MM channel and $\lambda$ is an overall coupling.
The radial wave functions $u_{c}$ and $v_{j}$ result from the separation of
the total wave function in spherical coordinates,
i.e., $\psi=\frac{u(r)}{r}Y_{lm}(\theta,\varphi)$. 

We now first solve the equation analytically for $r<a$ and $r>a$
ignoring the delta shell, and then match both solutions and their
derivatives at $r=a$, accounting for the delta function at this point.
For the quark-antiquark components, we introduce the parameter 
\begin{equation}
\nu_{c}=\frac{E}{\hbar\omega}-\frac{l_{c}}{2}-\frac{3}{4}
\end{equation}
and make the substitution
$u_{c}(r)=F_{c}(r)r^{1+L_{c}}e^{-\frac{1}{2}\mu_{c}\omega r^{2}}$.
This way we find that the solutions are of the form
\begin{equation}
u_{c}(r)=\left\{
\begin{array}{l}
a_{c}M(-\nu_{c},l_{c}+\frac{3}{2},\mu_{c}\omega r^{2})e^{-\frac{1}{2}\mu_{c}\omega r^{2}}r^{1+l_{c}}\;,\;\;r<a\\
b_{c}\:U(-\nu_{c},l_{c}+\frac{3}{2},\mu_{c}\omega r^{2})e^{-\frac{1}{2}\mu_{c}\omega r^{2}}r^{1+l_{c}}\;,\;\;r>a
\end{array} \right.
\label{eq:conf}
\end{equation}
where $M$ and $U$ are the Kummer and Tricomi confluent hypergeometric
functions (same as the $\Phi$ and $\Psi$ functions defined in Ref.~\cite{B53}
and employed in Ref.~\cite{EPJC73p2351}). Given the properties of these
functions, this guarantees that $u_{c}$ is a solution to
$\hat{h}_{q\bar{q}}^{c}u_{c}$ $=Eu_{c}$ for $r\neq a$,
regular at the origin, and vanishing at infinity.

For the two-meson wave function, we introduce the variable
$q_{j}=ip_{j}$ for each channel, with $p_{j}$ the corresponding relative
momentum. Then we have
\begin{equation}
E=M_{1}^{j}+M_{2}^{j}-\frac{\hbar^{2}}{2\mu_{j}}q_{j}^{2} \; .
\end{equation}
The two-meson components $v_{j}(r)$ can be written as
\begin{equation}
v_{j}(r)=\left\{
\begin{array}{l}
A_{j}i_{L_{j}}(q_{j}r)\, r \;,\;\; r<a \\
B_{j}k_{L_{j}}(q_{j}r)\, r \;,\;\; r>a
\end{array} \right.
\label{eq:scat}
\end{equation}
where $i_{l}(x)$ and $k_{l}(x)$ are the modified spherical
Bessel functions of the first and third kind, respectively 
(cf.\ the functions $J_l$ and $N_l$ in Ref.~\cite{EPJC73p2351}).
The function $i_{l}$ is regular at
the origin and divergent at infinity, whereas $k_{l}$ is irregular
at the origin and falls off exponentially as $x\rightarrow\infty$.
This solution is valid as long as the energy of the state is below
all two-meson thresholds.

For convenience, we now simplify our notation, by writing $M_{c}$ for
$M(-\nu_{c},l_{c}+\frac{3}{2},\mu_{c}\omega a^{2})$, $i_{j}$
for $i_{L_{j}}(q_{j}a)$, and similarly $U_c$ and $k_j$.
In order to determine the coefficients $a_{c}$, $b_{c}$, $A_{j}$, and
$B_{j}$, as well as the energy $E$ for a given coupling $\lambda$
or $\lambda$ for a given $E$, we first use continuity
of the wave function at $r=a$, i.e.,
\begin{eqnarray}
a_{c}M_{c}&=&b_{c}U_{c}\; , \label{eq:cont_conf} \\
A_{j}i_{j}&=&B_{j}k_{j} \; . \label{eq:cont_scatter}
\end{eqnarray}
Next we integrate the Schr\"{o}dinger equation (\ref{eq:schro}) from
$a-\epsilon$ to $a+\epsilon$, with $\epsilon$ infinitesimal. In doing so,
first for the $q\bar{q}$ components, we obtain
\begin{equation}
-\frac{\hbar^{2}}{2\mu_{c}}\big(u_{c}'(a^{+})-u_{c}'(a^{-})\big)+
\lambda\sum_{j}\frac{g_{cj}}{2\mu_{c}a}v_{j}(a)=0 \; .
\end{equation}
This yields, by substituting the expressions for $u_{c}$ and $v_{j}$
given in Eqs.~(\ref{eq:conf},\ref{eq:scat}),
\begin{equation}
b_{c}U_{c}'-a_{c}M_{c}'=\lambda\frac{e^{\frac{1}{2}\mu_{c}
\omega a^{2}}a^{-l-2}}{2\mu_{c}\omega}\sum_{j}g_{cj}A_{j}i_{j} \; .
\label{eq:der_conf}
\end{equation}
The same procedure applied to the MM channels gives the relations
\begin{equation}
-\frac{\hbar^{2}}{2\mu_{j}}\big(v_{j}'(a^{+})-v_{j}'(a^{-})\big)+
\lambda\sum_{c}\frac{g_{cj}}{2\mu_{c}a}u_{c}(a)=0 \;,
\end{equation}
and so using again Eqs.~(\ref{eq:conf},\ref{eq:scat}) we obtain
\begin{equation}
B_{j}k_{j}'-A_{j}i_{j}'=\lambda\sum_{c}\frac{g_{cj}\mu_{j}e^{-
\frac{1}{2}\mu_{c}\omega a^{2}}{}_{a}^{1+l}}{q_{j}\mu_{c}a^{2}}M_{c}a_{c} \; .
\label{eq:der_scatter}
\end{equation}
We can now eliminate $b_{c}$ and $B_{j}$ from
Eqs.~(\ref{eq:der_conf},\ref{eq:der_scatter}) by using
Eqs.~(\ref{eq:cont_conf},\ref{eq:cont_scatter}):
\begin{eqnarray}
b_{c}&=&\frac{M_{c}}{U_{c}}a_{c}\label{eq:bc_coef} \; , \\
B_{j}&=&\frac{i_{j}}{k_{j}}A_{j}\label{eq:Bcoef} \; .
\label{eq:bcbj}
\end{eqnarray}
Using next Eqs.~(\ref{eq:der_scatter},\ref{eq:Bcoef}), we can write
$A_{j}$ as a function of $a_{c}$, viz.\
\begin{equation}
\big(\frac{k_{j}'}{k_{j}}i_{j}-i_{j}'\big)\, A_{j}=\lambda\frac{g_{j}\mu_{j}
e^{-\frac{1}{2}m\omega a^{2}}a{}^{l-1}}{q_{j}\mu_{c}}M_{c}\, a_{c} \; .
\label{eq:Acoef}
\end{equation}
Substituting the latter result as well as Eq.~(\ref{eq:bc_coef}) in
Eq.~(\ref{eq:der_conf}), and defining $\alpha_{c}\equiv a_{c}M_{c}$, we find
\begin{align}
\Big(\frac{U_{c}'}{U_{c}}-\frac{M_{c}'}{M_{c}}\Big)\alpha_{c}= & \lambda^{2}\sum_{j}\frac{\mu_{j}g_{cj}}{2\mu_{c}\omega q_{j}a^{3}}\Big(\frac{k{}_{j}'}{k_{j}}-\frac{i_{j}'}{i_{j}}\Big)^{-1}\nonumber \\
 & \times \sum_{d}\frac{e^{\frac{1}{2}(\mu_{c}-\mu_{d})
\omega a^{2}}g_{dj}}{\mu_{d}}\,\alpha_{d} \; .
\end{align}
This set of equations is just a generalised eigensystem, with
$\lambda^{2}$ the generalised eigenvalue and $\alpha_{c}$ the generalised
eigenvector, which can be written as
\begin{equation}
D_{c}\,\alpha_{c}=\lambda^{2}G_{cd}\,\alpha_{d} \; ,
\label{eq:gev}
\end{equation}
where $D$ is a diagonal matrix.

We can use the latter equation to determine the value of $\lambda$
for which there is a certain bound state with a chosen energy $E$.
Alternatively, for a given $\lambda$, the energies of possible bound
state can be found by employing Newton's method to search for 
zeros of the determinant
\begin{equation}
\det(\mathrm{D}-\lambda^{2}\mathrm{G})=0 \; .
\end{equation}
Either way, the wave-function coefficients can next be calculated from
the obtained $\alpha_{c}$, using Eqs.~(\ref{eq:bc_coef}--\ref{eq:Acoef}).
Finally, the scale of the coefficients can be fixed by imposing the
normalisation condition
\begin{equation}
\sum_{c}\int_{0}^{\infty}dr\, u_{c}(r)^{2}+\sum_{j}\int_{0}^{\infty}dr\, v_{j}(r)^{2}=1 \; .
\end{equation}

\section{$\bm{X(3872)}$, $\bm{J/\psi}$, and $\bm{\psi(2S)}$ wave functions}
\label{charmonium}
Now that we have derived the general solution for the multichannel wave
functions resulting from Eq.~(\ref{eq:schro}), we should focus on the
specific wave functions of the axial-vector ($A$) charmonium system $X(3872)$
as well as the vector ($V$) states $\jpsi$ and $\psi(2S)$, since we shall
consider EM decays of the former charmonium into the latter two. In order to
account for all non-negligible meson-loop effects, we couple these systems to
OZI-allowed channels containing pairs of the ground-state
open-charm mesons $D$, $D^\star$, $D_s$ and $D_s^\star$, being either
pseudoscalar ($P$) or vector ($V$). Now, $V$ states can decay into the
combinations $PP$, $PV$, and $VV$, with odd orbital angular momentum $L$
because of parity conservation, whereas $A$ systems couple to channels
of the $PV$ and $VV$ types, with even $L$.
In Tables~\ref{psicoupl} and \ref{xcoupl} we give the relative couplings of
the different charmonia to the corresponding two-meson channels, viz.\
\begin{table}[h]
\protect\caption{Squares of coupling coefficients for the vector charmonia
$J/\psi$ and $\psi(2S)$. The generic notation $D$ represents $D^0$, $D^\pm$,
or $D_s^\pm$ open-charm mesons.}
\begin{center}
\begin{tabular}{ccccc}
\hline 
MM channels & L & S & $g_{(l=0)}$ & $g_{(l=2)}$\tabularnewline
\hline 
$DD$ & 1 & 0 & $1/24$ & $1/72$\tabularnewline
$DD^{*}$ & 1 & 1 & $1/6$ & $1/72$\tabularnewline
$D^{*}D^{*}$ & 1 & 0 & $1/72$ & $1/216$\tabularnewline
$D^{*}D^{*}$ & 1 & 2 & $5/18$ & $1/1080$\tabularnewline
$D^{*}D^{*}$ & 3 & 2 & 0 & $7/40$\tabularnewline
\hline 
\end{tabular}
\end{center}
\label{psicoupl}
\end{table}
\begin{table}[h]
\protect\caption{Squares of coupling coefficients for the axial-vector
charmonium $X(3872)$. Also see Table~\ref{psicoupl}.}
\begin{center}
\begin{tabular}{cccc}
\hline 
MM channels & L & S & $g_{(l=1)}$\tabularnewline
\hline 
$DD^{*}$ & 0 & 1 & $1/18$\tabularnewline
$DD^{*}$ & 2 & 1 & $5/72$\tabularnewline
$D^{*}D^{*}$ & 2 & 2 & $5/24$\tabularnewline
\hline 
\end{tabular}
\end{center}
\label{xcoupl}
\end{table}
for $\jpsi$ (or $\psi(2S)$) and $X(3872)$, respectively. These couplings
have been calculated employing the formalism of Ref.~\cite{ZPC21p291}, based on
overlaps in an HO basis. For economy, each
listed channel in these tables really represents three channels, with e.g.\
$DD$ standing for $D^0D^0$, $D^\pm D^\mp$, and $D_s^+D_s^-$, all with the same
coupling. Also note that in the $V$ case of $\jpsi$ and $\psi(2S)$ two
$c\bar{c}$ channels must be included, viz.\ \tso\ and \tdo, giving rise
to two sets of couplings.

Before we can compute the different wave functions, we must fix the model
parameters, viz.\ $\omega$ (HO frequency), $m_c$ (charm-quark mass), $\lambda$
(overall coupling constant), and $a$ (string-breaking distance). Now, the
former two parameters are not really free, as they have been kept fixed at the
values $\omega=190$~MeV and $m_c=1562$~MeV, determined in
Ref.~\cite{PRD27p1527}, in all subsequent work. Then, $\lambda$ and $a$ should
be adjusted to the masses of $\jpsi$, $\psi(2S)$, and $X(3872)$, which can be
done reasonably well, in spite of having only two parameters to fit three
observables. Nevertheless, we believe that in the present calculation it is
most important to have as accurate as possible wave functions, so that we
somewhat relax the usual condition of only one $\lambda$ for all described
systems. This way we are able to precisely reproduce the experimental
$\jpsi$, $\psi(2S)$, and $X(3872)$ masses, with the values
$\lambda_\psi=2.527$, $\lambda_X=2.176$, and $a=1.95$~GeV$^{-1}$, the coupling
$\lambda_\psi$ being of course the same for $\jpsi$ and $\psi(2S)$, as we have
included exactly the same MM channels for these two $V$ states. Again,
ideally there should be only one $\lambda$. However, one must realise that the
completeness property of the couplings $g_i$ as computed in the scheme of
Ref.~\cite{ZPC21p291}, which implies $\sum_i g_i^2=1$ for a system with any
quantum numbers, is only satisfied if {\em all} \/decay channels are included.
Well, in the latter HO-based formalism, the number of allowed channels
is finite but still huge and so too large to totally account for in
practical calculations. Therefore, when coupling only to the most important
channels, generally those with the lowest thresholds, somewhat different values
of $\lambda$ for clearly distinct systems are perfectly acceptable. And
indeed, $\lambda_\psi$ and $\lambda_X$ differ by only about 15\%. Moreover,
$\lambda_X=2.176$ is of the same order of magnitude as $\lambda\approx3$
obtained in the $X(3872)$ study of Ref.~\cite{EPJC71p1762}, in which the
related yet quantitatively different momentum-space RSE formalism was employed,
for a similar value of the decay radius, viz.\ $a=2$~GeV$^{-1}$.

\begin{figure}[ht]
\begin{tabular}{c}
\resizebox{!}{175pt}{\includegraphics{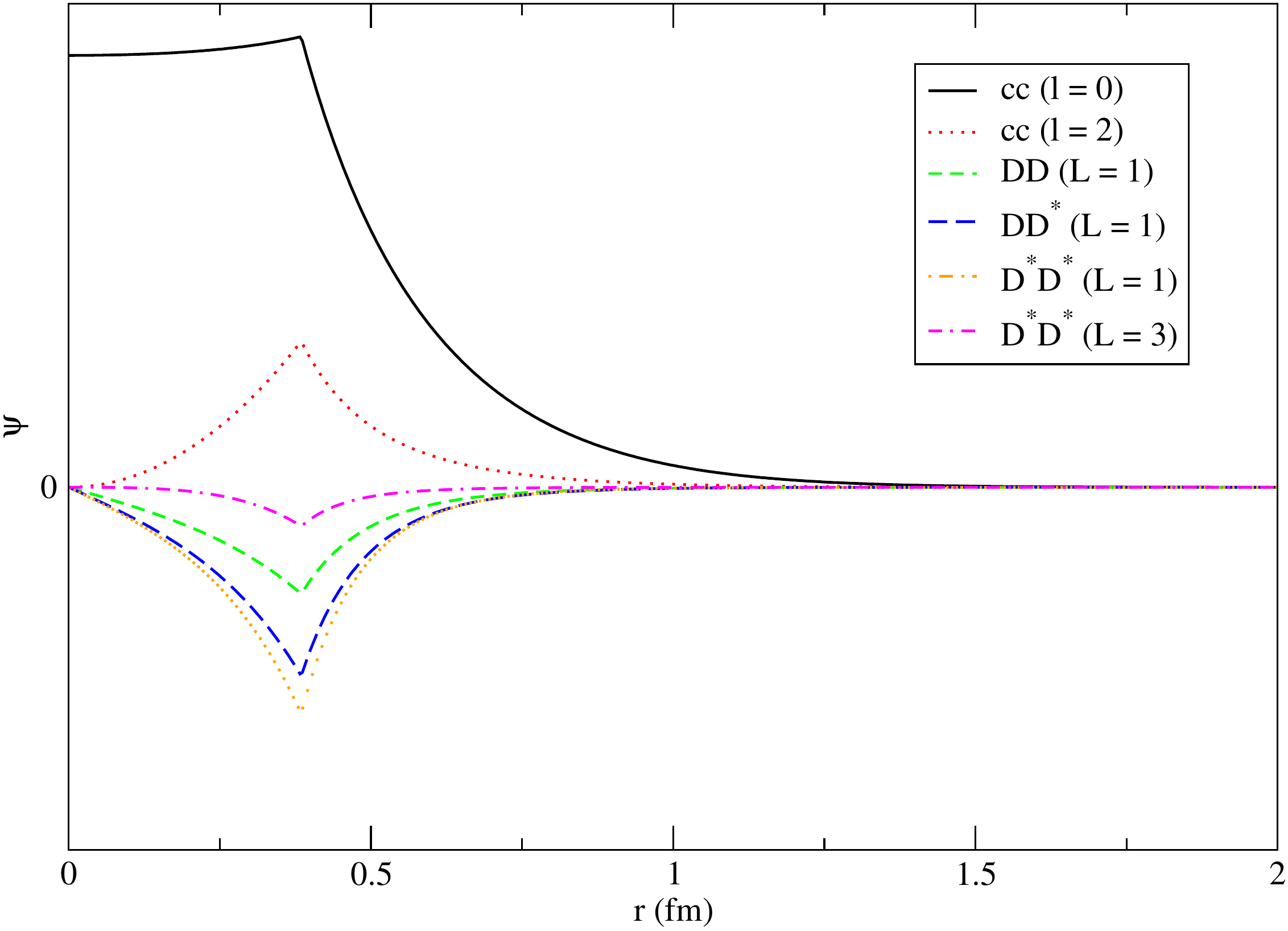}}\\[2mm]
\resizebox{!}{175pt}{\includegraphics{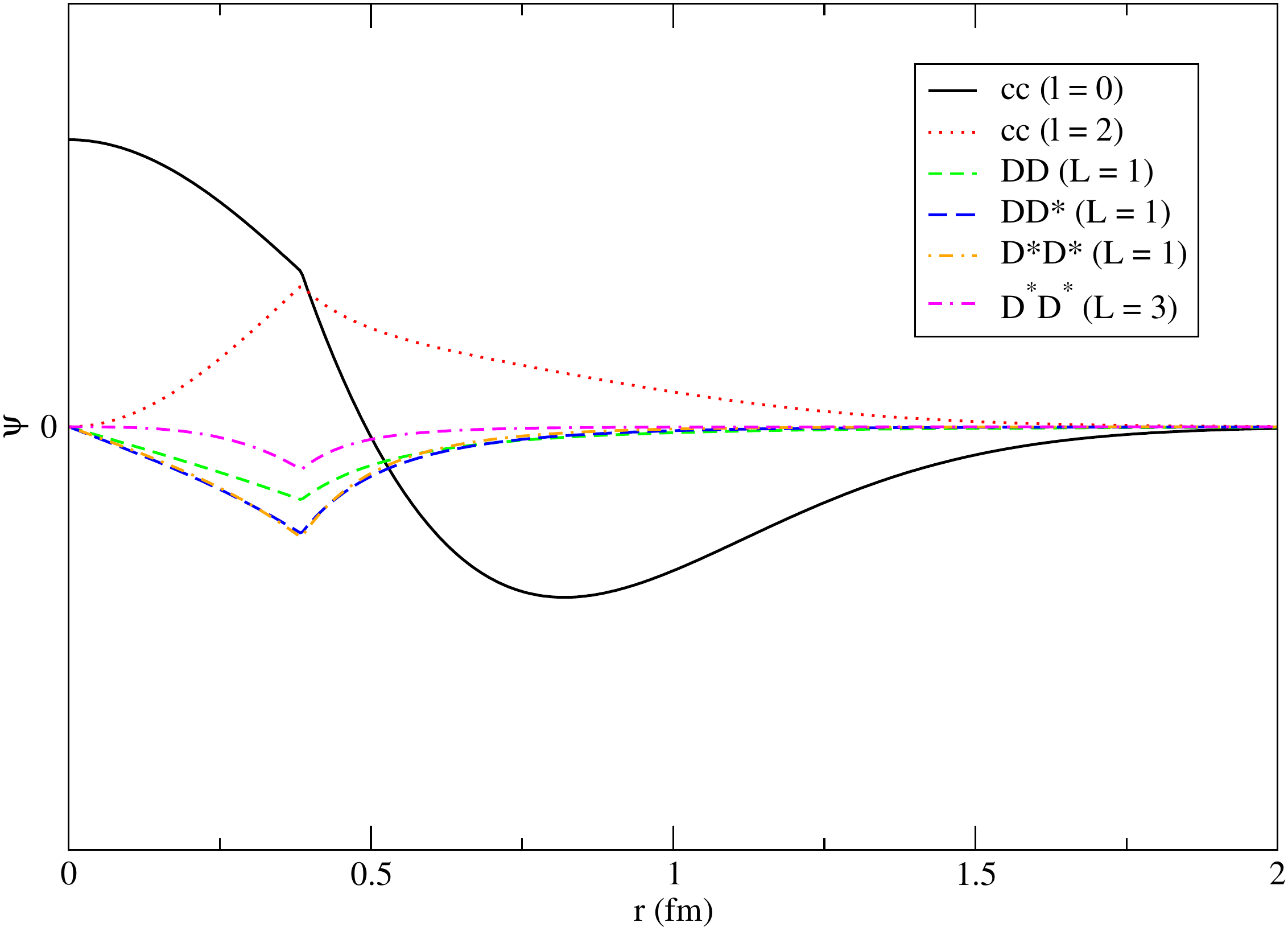}} 
\end{tabular}
\protect\caption{Multichannel components of $\jpsi$ (top) and $\psi(2S)$
(bottom) radial wave functions, in arbitrary units. Note that each MM
curve represents the r.m.s. value of three channel wave functions, with $DD$
standing for $D^0D^0$, $D^\pm D^\mp$, or $D_s^\pm D_s^\mp$, and so forth
(also see Table~\ref{psicoupl} and text). By convention and for clarity
purposes, the MM curves take negative values.}
\label{psiwf}
\end{figure}

\begin{figure}[!h]
\begin{tabular}{c}
\resizebox{!}{175pt}{\includegraphics{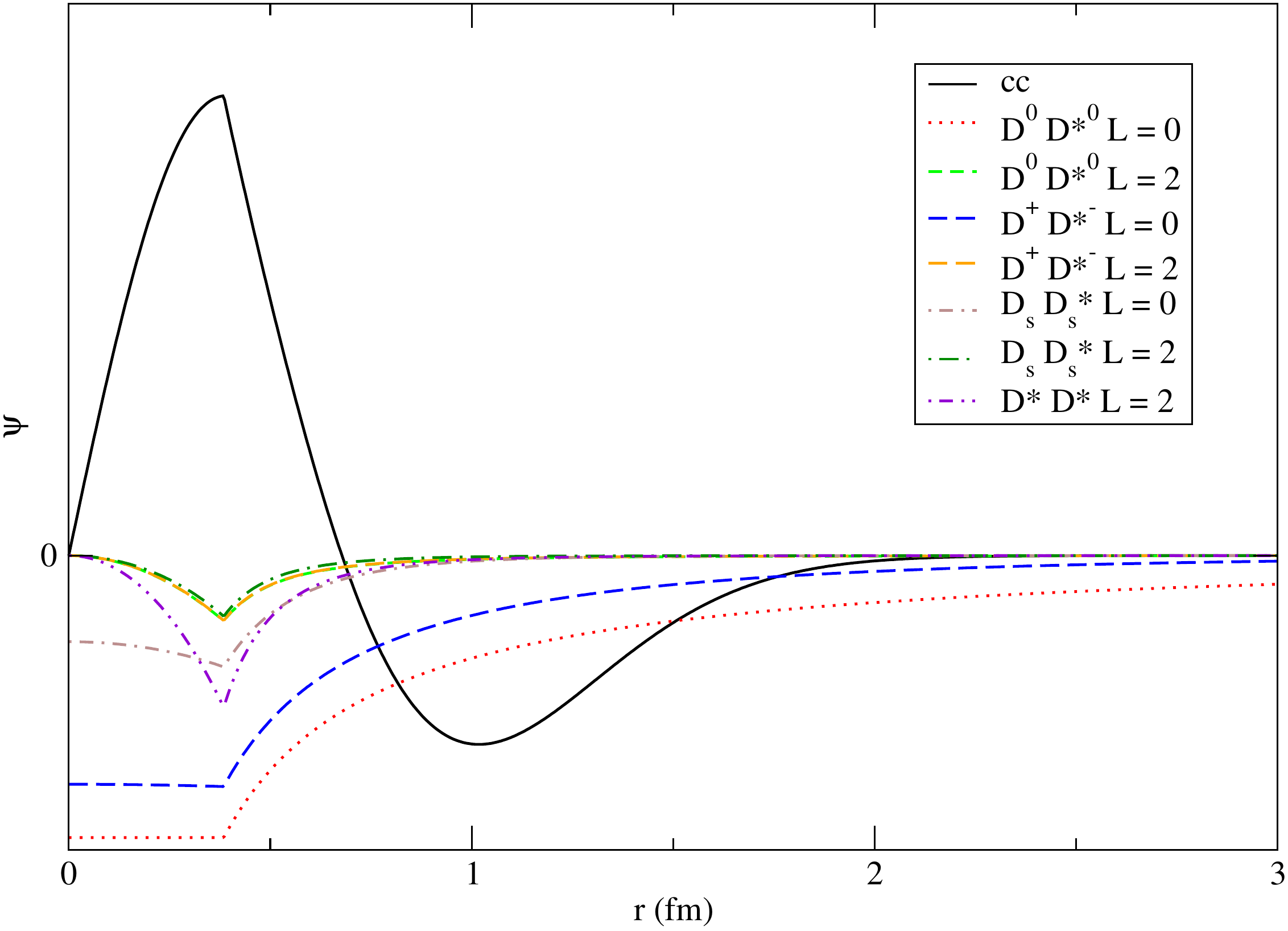}}
\end{tabular}
\protect\caption{Multichannel components of $X(3872)$ radial wave function,
in arbitrary units. Also see Fig.~\ref{psiwf} and Table~\ref{xcoupl}.}
\label{Xwf}
\end{figure}

Now we are in a position to compute the three needed radial wave functions,
using Eq.~(\ref{eq:gev}) and Eqs.~(\ref{eq:bc_coef}--\ref{eq:Acoef}).
In Fig.~\ref{psiwf} the $\jpsi$ and $\psi(2S)$ wave functions are displayed,
and in Fig.~\ref{Xwf}  that of $X(3872)$.
The first thing we observe in all three plots is a kink in the
wave-function components at $r=a$, which is a direct consequence of our
choosing a singular transition potential, mimicking string breaking at that
precise distance. Concerning the $V$ charmonia of Fig.~\ref{psiwf}, there is
a dominant $l=0$ $c\bar{c}$ wave function, which does not vanish at the origin,
but also considerable $l=2$ $c\bar{c}$ and $L=1$ MM components. In the case of
$X(3872)$, which has a seemingly dominant $l=1$ $c\bar{c}$ wave function, the
$L=0$ $D^\pm D^{\star\mp}$ and $D^0 D^{\star0}$ components are also very
sizable, especially the latter. As a matter of fact, the $D^0 D^{\star0}$
channel turns out to be the most important one in terms of probability, due to 
its very long tail, resulting from the small binding with respect to the
$X(3872)$ mass \cite{EPJC71p1762}. This and all other wave-function
probabilities are given in Tables~\ref{psiprob} and \ref{Xprob}, i.e., for
the $V$ charmonia and $X(3872)$, respectively.
\begin{table}[t]
\protect\caption{Percentage probabilities of $\jpsi$ and $\psi(2S)$
wave-function components, with the $c\bar{c}$ and MM numbers including 
$l=0,2$ and $L=1,3$ contributions, respectively. Also, $D$ stands generically
for $D^0$, $D^\pm$, or $D_s^\pm$, etc., as in Fig.~\ref{psiwf}.}
\begin{center}
\begin{tabular}{c|cccc}
\hline 
\% & $c\bar{c}$ & $DD$ & $DD^\star$ & $D^\star D^\star$ \tabularnewline
\hline 
$\jpsi$    & 83.63 & 2.07 & 6.02 & 8.28 \tabularnewline
$\psi(2S)$ & 94.50 & 1.29 & 2.08 & 2.12 \tabularnewline
\hline 
\end{tabular}
\end{center}
\label{psiprob}
\end{table}
What may look surprising in Table~\ref{psiprob} is that $\jpsi$ has clearly
larger MM components in its wave function than $\psi(2S)$. But this can be
understood by observing that the decay radius $a$ is relatively close to the
node in the $\psi(2S)$ $c\bar{c}$ wave function, which reduces the influence
of the MM channels. As for $X(3872)$, we see in Table~\ref{Xprob} that
\begin{table}[h]
\protect\caption{Percentage probabilities of $X(3872)$
wave-function components, with the MM figures including $L=0,2$ contributions.
Also, the $D^\star D^\star$ value accounts for the $D^{\star0}D^{\star0}$, 
$D^{\star\pm}D^{\star\mp}$, and $D_s^{\star\pm}D_s^{\star\mp}$ summed
contributions.}
\begin{center}
\begin{tabular}{c|ccccc}
\hline 
\% & $c\bar{c}$ & $D^0D^{\star0}$ & $D^\pm D^{\star\mp}$ & 
$D_s^\pm D_s^{\star\mp}$ & $D^\star D^\star$ \tabularnewline
\hline 
$X(3872)$  & 26.76 & 65.03 & 7.00 & 0.53 & 0.68 \tabularnewline
\hline 
\end{tabular}
\end{center}
\label{Xprob}
\end{table}
the $D^0 D^{\star0}$ component is by far the largest, just like in the
two-channel model study of Ref.~\cite{EPJC73p2351}. Nevertheless, we observe
a significant decrease in the latter MM channel's probability, and a large
increase in the $c\bar{c}$ probability, viz.\ from about 7.5\% to almost 27\%,
for a similar decay radius $a$. At first sight, this seems very surprising, as
in the present work we include several additional MM channels. However, it must
be realised that all such channels contribute to shift the $X(3872)$ bare
energy level of 3929~MeV down to the physical value of 3871.69~MeV, which 
reduces the relative importance of the $D^0D^{\star0}$ channel. Since it is
precisely the latter wave-function component that has a very large extension
in coordinate space, the reduction of its coupling will reduce its probability
roughly proportionally, mostly to the benefit of the $c\bar{c}$ component.

Another way to look at the different wave-function components is by focusing
on how the quenched solutions are modified by the coupling to the MM channels.
For that purpose, we compute the overlaps of the unquenched $c\bar{c}$
wave functions, which are the solid curves in Figs.~\ref{psiwf} (top,
bottom) and \ref{Xwf}, with pure HO solutions of different radial
quantum number $n$, the results being given in Tables~\ref{ovjpsi},
\ref{ovpsi2S}, and \ref{ovX}, respectively. Note that these numbers do
\begin{table}[ht]
\protect\caption{Overlap percentages of $\jpsi$ $c\bar{c}$ component with HO
functions.} 
\begin{center}
\begin{tabular}{c|cccc}
\hline 
\% & $n=0$ & $n=1$ & $n=2$ & $n=3$ \tabularnewline
\hline 
$l=0$ & 82.19 & 9.76 & 1.44 & 0.11 \tabularnewline
$l=2$ & 0.98 & 1.01 & 0.86 & 0.67 \tabularnewline
\hline 
\end{tabular}
\end{center}
\label{ovjpsi}
\end{table}
\begin{table}[h]
\protect\caption{Overlap percentages of $\psi(2S)$ $c\bar{c}$ component with
HO functions.}
\begin{center}
\begin{tabular}{c|cccc}
\hline 
\% & $n=0$ & $n=1$ & $n=2$ & $n=3$ \tabularnewline
\hline 
$l=0$ & 16.53 & 70.00 & 1.13 & 0.05 \tabularnewline
$l=2$ & 9.53 & 1.07 & 0.52 & 0.31 \tabularnewline
\hline 
\end{tabular}
\end{center}
\label{ovpsi2S}
\end{table}
\begin{table}[h]
\protect\caption{Overlap percentages of $X(3872)$ $c\bar{c}$ component with
HO functions.}
\begin{center}
\begin{tabular}{c|cccc}
\hline 
\% & $n=0$ & $n=1$ & $n=2$ & $n=3$ \tabularnewline
\hline 
$l=1$ & 8.82 & 85.37 & 4.14 & 1.05 \tabularnewline
\hline 
\end{tabular}
\end{center}
\label{ovX}
\end{table}
{\em not} \/concern the total wave-function overlaps, which can easily be
obtained through multiplication by the $c\bar{c}$ probabilities in
Tables~\ref{psiprob} and \ref{Xprob}. It is interesting to see that the
overlaps with the lowest four radial HO states, including of course also all
the $l$ states that couple, are almost sufficient to reconstruct the
$c\bar{c}$ wave functions, namely to 97\% for $\jpsi$, and to even more than
99\% for $\psi(2S)$ and $X(3872)$. This might be useful for quark models
with confinement mechanisms different from ours, in which sometimes HO wave
functions are used to compute certain observables. In the specific case of
$X(3872)$, we get an 8.82\% overlap with the \otpo\ bare HO state. This is
interesting, as in Ref.~\cite{PRD85p114002} such a component, despite being
smaller in size than our result, was found to have a large influence on the
EM decay widths of $X(3872)$.

To conclude this section about wave functions, we compute the r.m.s.\ radii of
$\jpsi$, $\psi(2S)$, and $X(3872)$, obtaining 0.456~fm, 0.930~fm, and 6.57~fm, 
respectively. Notice that the latter number is somewhat smaller than the value
found in Ref.~\cite{EPJC73p2351}, which is logical in view of the here reduced
influence of the long $D^0 D^{\star0}$ tail.

\section{Electromagnetic transitions}
\label{em}

In this section we review the formalism \cite{PRD44p2803}
for EM transitions of quarkonium systems coupled to MM channels. In order to
calculate the EM decay rate of a multicomponent meson state,
we couple the EM field to our coupled-channel strong-interaction
Hamiltonian $\hat{H}_{q\bar{q}-MM}$, obtaining a Hamiltonian of
the type
\begin{equation}
\hat{H}=\hat{H}_{q\bar{q}-MM}+\hat{H}_{em}+\hat{H}_{int} \; ,
\end{equation}
where $\hat{H}_{em}$ is the free EM part, in
Gaussian units reading
\begin{equation}
\hat{H}_{em}=\frac{1}{8\pi}\int d^{3}\mathbf{x}\,
(\mathbf{E}^{2}+\mathbf{B}^{2}) \;
\end{equation}
and $\hat{H}_{int}$ describes the interaction
between the hadrons and the EM field. This interaction
Hamiltonian can be naturally obtained from $\hat{H}_{q\bar{q}-MM}$ 
via a minimal-coupling prescription.
As we know, the hadronic coupled-channel Hamiltonian $\hat{H}_{q\bar{q}-MM}$
has diagonal elements of the form
\begin{equation}
\hat{h}=-\frac{\hbar^{2}}{2m_{1}}\nabla_{1}^{2}-\frac{\hbar^{2}}{2m_{2}}i
\nabla_{2}^{2}+V(\mathbf{x}_{1}-\mathbf{x}_{2}) \; .
\end{equation}
The Hamiltonian for hadrons interacting with radiation is then obtained
through the minimal coupling 
\begin{equation}
-\frac{\hbar^{2}}{2m}\nabla^{2}\rightarrow-\frac{\hbar^{2}}{2m}
(\nabla-\frac{iQ}{\hbar c}\mathbf{A})^{2} \; .
\label{eq:mincoupling}
\end{equation}
For the EM radiation field, we use the gauge conditions
\begin{equation}
\begin{array}{rcc}
A^{0} & = & 0 \; , \\
\nabla\cdot\mathbf{A} & = & 0 \; .
\label{eq:gaugecond}
\end{array}
\end{equation}
Using Eqs.~(\ref{eq:mincoupling}) and (\ref{eq:gaugecond}) and allowing
for a possible anomalous magnetic moment $\mu_{i}$, we get
\begin{equation}
\hat{h}_{\mbox{\scriptsize int}}=\sum_{i}\frac{iQ_{i}}{m_{i}c}\,\mathbf{A}
(\mathbf{x}_{i})\cdot \nabla_{i}-\mu_{i}\mathbf{S}_{i}\cdot\mathbf{B}
(\mathbf{x}_{i})+\!\frac{Q_{i}^{2}}{2m_{i}c^{2}}\mathbf{A}(\mathbf{x}_{i})^{2}
\,.
\end{equation}
Since we are considering the EM interaction perturbatively,
the term quadratic in $\mathbf{A}$ can and will be neglected.
Now, for quarks and antiquarks, the magnetic moment is given by
\begin{equation}
\mu_{i}=\frac{Q_{i}\hbar}{m_{i}c} \; .
\end{equation}
Then, the meson magnetic moment is obtained by assuming a pure
$q\bar{q}$ state, resulting in the expression
\begin{equation}
\mu_{i}=\frac{\hbar}{2c}\left(\frac{Q_{q}}{m_{q}}+
\frac{Q_{\bar{q}}}{m_{\bar{q}}}\right) \; .
\end{equation}
Note that, apart from accounting for the mesons' magnetic moments,
the present calculation neglects their internal structure. As a consequence,
to describe the EM decays of unitarised mesons, we shall only consider
processes of the type
\begin{equation}
(Q\bar{Q})^{*}\rightarrow Q\bar{Q}+\gamma
\end{equation}
and 
\begin{equation}
(M_{1}M_{2})^{*}\rightarrow M_{1}M_{2}+\gamma \; ,
\end{equation}
while neglecting the ones that change the internal structure of individual
mesons, viz.\
\begin{equation}
M_{1}^{*}M_{2}^{*}\rightarrow M_{1}M_{2}+\gamma \; .
\end{equation}

As the wave function of a unitarised meson has the form
\begin{equation}
|\psi_{\mbox{\scriptsize hadronic}}\rangle=\sum_{c}|\psi_{q\bar{q}}^{c}\rangle
+\sum_{j}|\psi_{MM}^{j}\rangle \; ,
\end{equation}
the total matrix element for an EM transition is given by
\begin{align}
\langle\Psi_{f}|\hat{H}_{int}|\Psi_{i}\rangle & =\sum_{cc'}\langle
\psi_{q\bar{q}}^{c}|\hat{h}_{int}^{cc'}|\psi_{q\bar{q}}^{c'}\rangle+
\sum_{jj'}\langle\psi_{MM}^{j}|\hat{h}_{int}^{jj'}|\psi_{MM}^{j'}\rangle\; .
\end{align}
The quantised EM vector potential $\mathbf{A}$ can be written in Gaussian
units as \cite{PRD44p2803}
\begin{align}
\mathbf{A}(\mathbf{r},t)\!=\!\sqrt{4\pi}\hbar c\sum_{\lambda lm}\int\!\frac{dk}
{2\pi}\frac{1}{\sqrt{2\omega_{k}}}\big[ & \mathbf{f}_{klm}^{(\lambda)}
(\mathbf{r})e^{-i\omega_{k}t}a_{\lambda lm}(k)+\nonumber \\
& \!\!\mathbf{f}_{klm}^{(\lambda)}(\mathbf{r})^{*}e^{-i\omega_{k}t}
a_{\lambda lm}^{\dagger}(k)\big].
\label{eq:vector_potential}
\end{align}
where the index $\lambda\in\{e,m\}$ indicates whether the component
is an electric multipole or a magnetic multipole. For instance,
the components with $\lambda=e$ and $l=2$ correspond to electric
quadrupole (E2) radiation, while those with $\lambda=m$ and $l=1$
stand for magnetic dipole (M1) radiation. Furthermore, $a_{\lambda lm}(k)$ and
$a_{\lambda lm}^{\dagger}(k)$ are photon annihilation and creation
operators obeying the commutation relations
\begin{equation}
[a_{\lambda lm}(k),a_{\lambda'l'm'}^{\dagger}(k')]=2\pi\delta(k'-k)
\delta_{\lambda'\lambda}\delta_{l'l}\delta_{m'm} \; .
\end{equation}
The vector fields $\mathbf{f}_{kJM}^{(\lambda)}(\mathbf{r})$ are
given by \cite{PRD44p2803}
\begin{equation}
\mathbf{f}_{klm}^{(m)}(\mathbf{r})=2kj_{l}(kr)\sum_{q}C_{lm-q1q}^{lm}Y_{lm-q}(\theta,\varphi)\hat{\mathbf{e}}_{q}\label{eq:funcm}
\end{equation}
and
\begin{align}
\mathbf{f}_{klm}^{(e)}(\mathbf{r})= & 2k\sum_{q}\big[\nonumber \\
 & j_{l-1}(kr)\sqrt{\frac{l+1}{2l+1}}C_{l-1,m-q,1q}^{l-1,m}Y_{l-1,m-q}(\hat{\mathbf{r}})\nonumber \\
 & -j_{l+1}(kr)\sqrt{\frac{l}{2l+1}}C_{l+1,m-q,1q}^{l+1,m}Y_{l+1,m-q}(\hat{\mathbf{r}})\nonumber \\
 & \big]\hat{\mathbf{e}}_{q} \; ,
\label{eq:funce}
\end{align}
with
\begin{eqnarray*}
\hat{\mathbf{e}}_{\pm1} & = & \mp\frac{\hat{\mathbf{e}}_{x}\pm 
i\hat{\mathbf{e}}_{y}}{\sqrt{2}} \; , \\
\hat{\mathbf{e}}_{0} & = & \hat{\mathbf{e}}_{z} \; .
\end{eqnarray*}
They have the properties \cite{PRD44p2803} 
\begin{eqnarray}
\nabla\times\mathbf{f}_{klm}^{(e)} & = & -ik\mathbf{f}_{klm}^{(m)}\;,\\
\nabla\times\mathbf{f}_{klm}^{(m)} & = & ik\mathbf{f}_{klm}^{(e)}\;,\\
\nabla.\mathbf{f}_{klm}^{(\lambda)} & = & 0\;,\\
\int d^{3}\mathbf{x}\,\mathbf{f}_{klm}^{(\lambda)}(\mathbf{x})\cdot
\mathbf{f}_{k'l'm'}^{(\lambda')} & = & 2\pi\delta(k'-k)\delta_{ll'}
\delta_{mm'}\delta_{\lambda\lambda'} \; .
\end{eqnarray}
Now, we have an initial state $|\Psi_{i}\rangle=|nJM\rangle\otimes|0\rangle$,
where $|0\rangle$ denotes the photon vacuum, and a final state 
$|\Psi_{f}\rangle=|n'J'M'\rangle\otimes|\lambda klm\rangle$,
where
\begin{equation}
|\lambda klm\rangle=a_{\lambda klm}^{\dagger}|0\rangle
\end{equation}
Substituting next Eqs.~(\ref{eq:vector_potential}), (\ref{eq:funcm}), and
(\ref{eq:funce}) into the matrix element 
\begin{equation}
\mathcal{M}_{if}=\langle\Psi_{f}|\hat{H}_{int}|\Psi_{i}\rangle \; ,
\end{equation}
we finally obtain, after a laborious yet straightforward calculation, 
the electric and magnetic decay matrix elements \cite{PRD44p2803}.
The results are given in Appendix~\ref{electromagnetic}.

\section{Results and comparison of EM decay rates}
\label{results}

Now we can present the results for our model calculation of the $X(3872)$
EM decays. First, though, we show in Table~\ref{expem} the up-to-date 
\begin{table}[ht]
\protect\caption{Experimental results for EM decays of $X(3872)$.}
\begin{tabular}{c|ccc}
\hline 
Experiment & $\mathcal{B}_{\gamma\jpsi}$ & $\mathcal{B}_{\gamma\psi(2S)}$ &
$\mathcal{R}_{\gamma\psi}$ \tabularnewline
\hline 
BaBar \cite{PRL102p132001} & $>9\times10^{-3}$ & $>0.030$ & $3.4\pm1.4$ 
\tabularnewline
Belle \cite{PRL107p091803} & $>6\times10^{-3}$ & \mbox{not seen} & $<2.1$
\tabularnewline
LHCb \cite{NPB886p665} & - & - & $2.46\pm0.64\pm0.29$ \tabularnewline
\hline 
\end{tabular}
\label{expem}
\end{table}
experimental status of such decays, which is clearly poor. First of all,
only lower bounds are reported for the $\gamma\jpsi$ and $\gamma\psi(2S)$
rates. On the other hand, for the very small total $X(3872)$ width, merely an
upper bound of 1.2~MeV is listed \cite{PDG2014}. This is understandable, as
small enough bin sizes to pin down the width more precisely are not possible
with present-day statistics. But as a consequence, the absolute magnitudes
of the two observed EM decays are largely unknown. Only their ratio 
\begin{equation}
\mathcal{R}_{\gamma\psi}\equiv\frac{\Gamma\left(X(3872)\rightarrow
\gamma\psi(2S)\right)}{\Gamma\left(X(3872)\rightarrow\gamma J/\psi\right)} 
\label{rgjgpsi}
\end{equation}
has been determined by two experiments, though still with large errors (see
Table~\ref{expem}). Coming now to our model predictions, we first observe that
a process  of the type $^{3\!}P_1\to\gamma\,{}^{3\!}S_1/{}^{3\!}D_1$ is
dominated by an electric dipole (E1) transition, besides a smaller 
magnetic quadrupole (M2) contribution. Using the expressions in
Appendix~\ref{electromagnetic}, we then obtain the results presented in 
Table~\ref{emresults}.
\begin{table}[h]
\protect\caption{Predictions (in keV) of E1 and M2 EM decays widths.}
\begin{center}
\begin{tabular}{c|cc}
\hline 
Process & E1 & M2 \tabularnewline
\hline 
$X(3872)\to\gamma\jpsi$    & 24.2 & 0.44 \tabularnewline
$X(3872)\to\gamma\psi(2S)$ & 28.8 & 0.07 \tabularnewline
\hline 
\end{tabular}
\end{center}
\label{emresults}
\end{table}
As expected, the M2 widths are much smaller than the E1 ones, since higher
multipoles are  suppressed by powers of photon momentum divided by (charm)
quark mass \cite{PRD78p114011}. Such a behaviour is roughly confirmed by
our numbers, as the photon in the process with $\jpsi$ in the final state has
about four times as much momentum as in the decay to $\psi(2S)$ \cite{PDG2014}.
Clearly, experimental statistics is insufficient so far to do an angular
analysis needed \cite{PRD84p092006} for disentangling the E1 and M2
contributions in the $X(3872)$ EM data. Therefore, in order to compare our
prediction for the EM branching-rate ratio defined in Eq.~(\ref{rgjgpsi}) with
the experimental values given in Table~\ref{expem}, we simply sum the E1 and M2
contributions, obtaining $\mathcal{R}_{\gamma\psi}=1.17$. This number is
compatible with the Belle \cite{PRL107p091803} upper bound of 2.1, but
somewhat too small as compared to the BaBar \cite{PRL102p132001} and LHCb
\cite{NPB886p665} measurements. However, the experimental values are only
marginally in agreement with one another and the uncertainties are still very
large. If we take our absolute EM-width predictions at face value, in
particular the result 28.9~keV for
$\Gamma\left(X(3872)\rightarrow\gamma\psi(2S)\right)$, the experimental lower
bound $\mathcal{B}_{\gamma\psi(2S)}>0.030$ \cite{PRL102p132001} implies
$\Gamma\left(X(3872)\right)$ $<1$~MeV, slightly lower than the PDG
\cite{PDG2014} upper limit of 1.2~MeV. Nevertheless, the $X(3872)$ total width
is probably even smaller than that, considering the $S$-matrix pole
trajectories of the $X(3872)$ resonance near the $D^0D^{\star0}$
threshold in the multichannel RSE calculation of Ref.~\cite{EPJC71p1762}.

Next, in Table~\ref{emcomp} we compare the present EM results to those of a
\begin{table}[t]
\protect\caption{Comparison of model predictions for the EM decays of
$X(3872)$. Abbreviations: mol = molecule, unq = unquenched $c\bar{c}$, EFT=
effective field theory, @ = present work. Models assume $1^{++}$ quantum
numbers, unless  otherwise specified.}
\begin{center}
\begin{tabular}{c|cccc}
\hline 
Ref.\ & $\Gamma_{\gamma\jpsi}\,\mbox{(keV)}$ & 
$\Gamma_{\gamma\psi(2S)}\,\mbox{(keV)}$ &
$\mathcal{R}_{\gamma\psi}$ & Notes\tabularnewline
\hline 
\cite{PRD86p113007} & 117 & - & - & mol\ \tabularnewline
\cite{PRD85p114002} & 76.6 & 62.8 & $0.8\pm0.2$ & unq \tabularnewline
\cite{PRD69p054008} & 11.0 & 63.9 & 5.81 & $c\bar{c}$\tabularnewline
\cite{PRD77p094013} & 124.8--251.4 & - & - & mol \tabularnewline
\cite{JPG38p015001} & 1.94--16.8 & 6.8--58.8 & 3.5 & $c\bar{c}$+mol
\tabularnewline
\cite{PRD84p014006} & $\sim10$ & - & - & $c\bar{c}q\bar{q}$ \tabularnewline
\cite{PTP126p91} & 1.86--2.09 & 5.06--6.54 & 2.43--3.52 & mol/$2^{-+}$\tabularnewline
\cite{PRD84p114026} & $3.54$ & $0.365$ & 0.10 & $c\bar{c}$/$2^{-+}$\tabularnewline
\cite{PRD83p094009} & - & $>70$ & - & EFT \tabularnewline
\cite{PLB598p197} & 71--139 & 94--95 & 0.68--1.34 & $c\bar{c}$\tabularnewline
\cite{PLB598p197} & 8 & 0.03 & $4\times10^{-3}$ & mol \tabularnewline
\cite{PLB697p233} & 33 & 146 & 4.4 & $c\bar{c}$ \tabularnewline
\cite{JPG40p035003} & 1.59 & 0.0029 & 0.002 & $c\bar{c}$/$2^{-+}$
\tabularnewline
\cite{PRD82p116002} & $\simeq1.8\times10^3$ & - & - &
$c\bar{c}$+$c\bar{q}q\bar{c}$ \tabularnewline
\hline 
@ & 24.7 & 28.9 & 1.17 & unq \tabularnewline
\hline 
\end{tabular}
\end{center}
\label{emcomp}
\end{table}
number of other model calculations (also see Ref.~\cite{ARXIV14107729}).
Clearly, our values are somewhere in the
middle of the ballpark of often disparate numbers. Generally, one may conclude
that the experimental EM rate ratio seems to favour models based on a \ttpo\
$c\bar{c}$ assignment for $X(3872)$, with or without other components.

Finally, we have a look at the importance of unquenching in our model. To that
end, we compute the EM predictions from bare HO $c\bar{c}$ wave functions
only, with unchanged parameters except for the overall couplings to MM
channels, which are set to zero. Note that this results in bare $\jpsi$,
$\psi(2S)$, and $X(3872)$ masses that are roughly 300, 100, and 100~MeV larger
than the physical ones, respectively. Having this proviso in mind, we obtain
for the total EM widths $\Gamma_{\mbox{\scriptsize quenched}}
\left(X(3872)\to\gamma\jpsi\right)=0.61$ keV and
$\Gamma_{\mbox{\scriptsize quenched}}\left(X(3872)\to\gamma\psi(2S)\right)=159$
~keV. Here we should remark that the former width, which corresponds to an
$2P\to1S$ EM transition, is sometimes reported \cite{PRD46p3862,ARXIV14107729}
to vanish identically for three-dimensional HO wave functions. However, this
is only true in the long-wave-length (alias dipole) approximation,
used in e.g.\ Ref.~\cite{PRD46p3862}, and not in the more general formalism
of Ref.~\cite{PRD44p2803} employed here. The latter article showed that the
dipole approximation is rather poor for several E1 transitions in charmonium.
Nevertheless, the very small quenched EM width we find for
$X(3872)\to\gamma\jpsi$ shows that in this particular situation the
approximation looks reasonable.  Anyhow, comparing to the unquenched total
(E1+M2) widths in Table~\ref{emresults}, we see a rather dramatic importance
of unquenching.

Concluding our study of multichannel effects, we calculate the separate
contributions from the $c\bar{c}$ and the MM channels. Note that this is done
in the full model and so using the wave functions plotted in Figs.~\ref{psiwf}
and \ref{Xwf}. The total exclusive $c\bar{c}$ and MM results are then obtained
by setting the wave functions of the MM and $c\bar{c}$ components to zero by
hand, respectively. Thus, we get $\Gamma^{c\bar{c}}_{\gamma\jpsi}=15.3$~keV,
$\Gamma^{MM}_{\gamma\jpsi}=1.12$~keV,
$\Gamma^{c\bar{c}}_{\gamma\psi(2S)}=28.0$~keV, and
$\Gamma^{MM}_{\gamma\psi(2S)}=0.01$~keV. These results reveal constructive
interference effects, especially in the $\jpsi$ case, as one cannot just sum
the $c\bar{c}$ and MM numbers to obtain the widths in Table~\ref{emresults}.

\section{Summary and conclusions}
\label{conclusions}
This paper is the third part of a triptych aimed at understanding the
microscopic dynamics of the enigmatic charmonium state $X(3872)$. The first
article \cite{EPJC71p1762} was devoted to its hadronic decays, including the
OZI-forbidden ones, arriving at a good description of the available data. The
second paper \cite{EPJC73p2351} focused on the importance of $X(3872)$'s
$c\bar{c}$ component, concluding that a purely molecular assignment is ruled
out. In our present work, we have generalised the $r$-space method employed in
the latter paper so as to include all virtual open-charm MM channels that may 
contribute significantly to the $X(3872)$ wave function, as well as to those
of the vector charmonia $\jpsi$ and $\psi(2S)$. These wave functions have then
been used to compute the widths of the experimentally observed EM decay
processes $X(3872)\rightarrow\gamma J/\psi$ and
$X(3872)\rightarrow\gamma\psi(2S)$, employing the formalism developed and 
applied to charmonium and bottomonium in Ref.~\cite{PRD44p2803}.

In the first place, and concerning the thus obtained wave functions, it is
quite remarkable that the inclusion of several additional MM channels to
describe $X(3872)$, most notably the $D^\pm D^{\star\mp}$ channel bound by only
about 8 MeV, gives rise to an increase of the $c\bar{c}$ probability from
7.48\% \cite{EPJC73p2351} to 26.76\%. Accordingly, the $c\bar{c}$
wave-function component becomes clearly the dominant one, except at very small
and very large distances. However surprising at first sight, this effect can be
explained by the reduced influence of the narrowly bound $D^0D^{\star0}$
channel, which has an extremely long tail and so takes up most of the total
probability. This is also confirmed by the reduction of the $X(3872)$ r.m.s.\
radius from 7.82~fm in Ref.~\cite{EPJC73p2351} to 6.57~fm here. As for the
$\jpsi$ and $\psi(2S)$ wave functions, we find significant \tdo\ $c\bar{c}$
and $P$-wave MM components, the former most prominently for $\psi(2S)$ and the
latter especially in the $\jpsi$ case. The closeness of the $\psi(2S)$
wave-function node to the decay radius $r$ offers an explanation for the
relatively reduced importance of MM channels for the first radially excited
vector charmonium state.

Coming now to the EM transitions of $X(3872)$, we obtain total widths of 
28.9~keV and 24.7~keV for the decays to $\gamma\psi(2S)$ and $\gamma\jpsi$,
respectively, with rate ratio $\mathcal{R}_{\gamma\psi}=1.17$. While there
are no data for the absolute magnitudes of the EM widths, the
latter ratio can be compared to three experimental values. Thus, our prediction
of 1.17 is fully compatible with the upper bound of 2.1 observed by the Belle
Collaboration \cite{PRL107p091803}, but a little bit too low for the numbers
$3.4\pm1.4$ and $2.46\pm0.64\pm0.29$ reported by BaBar \cite{PRL102p132001} and
LHCb \cite{NPB886p665}, respectively.
Although the experimental error bars are still quite large and the three
observations so far are nonetheless only in marginal agreement with one
another, additional mechanisms --- not considered here --- might remove the
slight discrepancy. One possibility is the inclusion of photonic decays of
individual mesons in the MM channels, that is, observed \cite{PDG2014}
processes of the type $D^{\star+}\to D^+\gamma$, $D^{\star0}\to D^0\gamma$, and
$D_s^{\star+}\to D_s^+\gamma$. Another source may be relativity, as
relativistic effects are capable of shifting the nodes in the $X(3872)$ and
$\psi(2S)$ $c\bar{c}$ wave-function components \cite{PRD44p2803}, thus
affecting the overlap integrals. While these issues are certainly worthwhile
to be studied in future work, the most important contribution to an even better
understanding of $X(3872)$ will be improved measurements, with higher
statistics and smaller bin sizes ($<1$~MeV if ever possible), in order to pin
down the absolute magnitudes of the EM decay widths, as well as the total
width.
\begin{acknowledgement}
This work was partially supported by
the \emph{Funda\c{c}\~{a}o para a Ci\^{e}ncia e a Tecnologia}
\/of the \emph{Mi\-nist\'{e}rio da Educa\c{c}\~{a}o e Ci\^{e}ncia}
\/of Portugal, through fellowship no.\ SFRH/BPD/73140/2010.
\end{acknowledgement}

\onecolumn 

\appendix
\section{EM transition matrix elements}
\label{electromagnetic}

The multipolar electric transitions matrix elements are given by
\cite{PRD44p2803}
\begin{align}
\mathcal{M}_{if}^{e}= & i\sqrt{\hbar\omega}(-1)^{m+L'}C_{M\, m\, M'}^
{J\, l\, J'}\sqrt{(2J+1)(2L+1)(2l+1)(2L'+1)}\left(\begin{array}{ccc}
L' & l & L\\
0 & 0 & 0
\end{array}\right)\Bigg[\nonumber \\
 & Q(-1)^{J+L'+S'+1}\frac{\delta_{SS'}}{\sqrt{l(l+1)}}\left\{ 
\begin{array}{ccc}
J & l & J'\\
L' & S' & L
\end{array}\right\} (R_{1}+R_{2})\nonumber \\
 & +(-1)^{1+S_{1}+S_{2}}\sqrt{(2l+1)(2S+1)(2S'+1)}\left\{ 
\begin{array}{ccc}
L & l & J\\
S & 1 & S'\\
J & L & J'
\end{array}\right\} \nonumber \\
 & \Big[(-1)^{S}2\mu_{1}\frac{\hbar\omega}{c}\sqrt{S_{1}(S_{1}+1)(2S_{1}+1)}
\left\{ \begin{array}{ccc}
S & 1 & S'\\
S_{1} & S_{2} & S_{1}
\end{array}\right\} R_{0,l}^{(1)}-\nonumber \\
 & (-1)^{l+S'}2\mu_{2}\frac{\hbar\omega}{c}\sqrt{S_{2}(S_{2}+1)(2S_{2}+1)}
\left\{ \begin{array}{ccc}
S & 1 & S'\\
S_{2} & S_{1} & S_{2}
\end{array}\right\} R_{0,l}^{(2)}\Big]\Bigg] \; ,
\end{align}
with the radial integrals
\begin{eqnarray}
R_{0,l}^{(i)} & = & \int_{0}^{\infty}dr\, u_{f}(r)j_{l}
\left(\frac{\mu}{m_{i}}kr\right) u_{i}(r) \; ,\\
R_{1} & = & \int_{0}^{\infty}dr\, u_{f}(r)\left[1+r\frac{\partial}{\partial r}
\right]\left[j_{l}\left(\frac{\mu}{m_{1}}kr\right)-(-1)^{l}j_{l}\left(
\frac{\mu}{m_{2}}kr\right)\right]u_{i}(r)\;,\\
R_{2} & = & \int_{0}^{\infty}dr\, u_{f}(r)\left[\frac{\mu}{m_{1}}
\frac{\hbar\omega}{m_{1}c^{2}}\,j_{l}\left(\frac{\mu}{m_{1}}kr\right)-(-1)^{l}
\frac{\mu}{m_{2}}\frac{\hbar\omega}{m_{2}c^{2}}\,j_{l}\left(
\frac{\mu}{m_{2}}kr\right)\right]\,
r\frac{\partial}{\partial r}u_{i}(r) \; . \\ \nonumber
\end{eqnarray}
The multipolar magnetic transitions matrix elements are \cite{PRD44p2803}
\begin{align*}
\mathcal{M}_{if}^{m}= \,& i\sqrt{\hbar\omega}(-1)^{m+L'+1}
C_{M\, m\, M'}^{J\, l\, J'}\sqrt{(2J+1)(2L'+1)}\left(\begin{array}{ccc}
L' & l & L\\
0 & 0 & 0
\end{array}\right)\Bigg[\\
 & Q\delta_{SS'}(-1)^{J+L'+S'}(2l+1)\left\{ \begin{array}{ccc}
L' & L & l\\
J & J' & S
\end{array}\right\} \\
 & \Big[\sqrt{(2L+3)(L+1)}\left(\begin{array}{ccc}
L' & l & L+1\\
0 & 0 & 0
\end{array}\right)\left\{ \begin{array}{ccc}
L' & L & l\\
1 & l & L+1
\end{array}\right\} R_{3,-L}\\
 & -\sqrt{(2L-1)L}\left(\begin{array}{ccc}
L' & l & L-1\\
0 & 0 & 0
\end{array}\right)\left\{ \begin{array}{ccc}
L' & L & l\\
1 & l & L-1
\end{array}\right\} R_{3,L+1}\Big]\\
 & 2(-1)^{S_{1}+S_{2}+1}\sqrt{(2L+1)(2S+1)(2S'+1)}\\
 & \bigg[\mu_{1}\frac{\hbar\omega}{c}(-1)^{S}\sqrt{S_{1}(S_{1}+1)(2S_{1}+1)}\left\{ \begin{array}{ccc}
S & 1 & S'\\
S_{1} & S_{2} & S_{1}
\end{array}\right\} \\
 & \Big[\sqrt{l(2l+3)}\left\{ \begin{array}{ccc}
L & l+1 & L'\\
S & 1 & S'\\
J & l & J'
\end{array}\right\} \left(\begin{array}{ccc}
L' & l+1 & L\\
0 & 0 & 0
\end{array}\right)R_{0,l+1}^{(1)}-\\
 & \sqrt{(l+1)(2l-1)}\left\{ \begin{array}{ccc}
L & l-1 & L'\\
S & 1 & S'\\
J & l & J'
\end{array}\right\} \left(\begin{array}{ccc}
L' & l-1 & L\\
0 & 0 & 0
\end{array}\right)R_{0,l-1}^{(1)}\Big]\\
 & +(-1)^{l+S'}\mu_{2}\frac{\hbar\omega}{c}\sqrt{S_{2}(S_{2}+1)(2S_{2}+1)}\left\{ \begin{array}{ccc}
S & 1 & S'\\
S_{2} & S_{1} & S_{2}
\end{array}\right\} \\
 & \Big[\sqrt{l(2l+3)}\left\{ \begin{array}{ccc}
L & l+1 & L'\\
S & 1 & S'\\
J & l & J'
\end{array}\right\} \left(\begin{array}{ccc}
L' & l+1 & L\\
0 & 0 & 0
\end{array}\right)R_{0,l+1}^{(2)}-\\
 & \sqrt{(l+1)(2l-1)}\left\{ \begin{array}{ccc}
L & l-1 & L'\\
S & 1 & S'\\
J & l & J'
\end{array}\right\} \left(\begin{array}{ccc}
L' & l-1 & L\\
0 & 0 & 0
\end{array}\right)R_{0,l-1}^{(2)}\Big]\bigg]\Bigg] \; ,
\end{align*}
with 
\begin{eqnarray*}
R_{3L} & = & \int_{0}^{\infty}dr\, u_{f}(r)\left[\frac{\hbar}{m_{1}c}\,
j_{l}\left(\frac{\mu}{m_{1}}kr\right)-(-1)^{l}\frac{\hbar}{m_{2}c}\,
j_{l}\left(\frac{\mu}{m_{2}}kr\right)\right]\,\left[
\frac{\partial}{\partial r}+\frac{L}{r}\right]\,u_{i}(r) \; .
\end{eqnarray*}
Accounting for the recoil of the final-state meson, the photon momentum
$k$ is given by
\begin{equation}
k=\frac{M_{i}^{2}-M_{f}^{2}}{2M_{i}}c \; .
\end{equation}
The decay width is then given by the Fermi Golden Rule
\begin{equation}
\Gamma^{(\lambda)}=\frac{2\pi}{\hbar}|\mathcal{M}_{if}^{\lambda}|^{2}\rho_{f}
\; ,
\end{equation}
where $\rho_{f}$ is the density of final states $\rho_{f}=1/2\pi\hbar c$.
\end{document}